\documentclass[reprint,twocolumn,superscriptaddress,preprintnumbers,amsmath,amssymb,aps,prl,floatfix]{revtex4-2}
\usepackage[utf8]{inputenc}

\usepackage{mathtools}
\usepackage{amsfonts}
\usepackage{mathrsfs}
\usepackage{bm}
\usepackage[normalem]{ulem}
\usepackage{bbold}
\usepackage{braket}
\usepackage{slashed}
\usepackage{dsfont}
\usepackage{dcolumn}%
\usepackage{amssymb,amsmath}
\usepackage{soul}  %

\usepackage{graphicx}
\usepackage{array}

\usepackage{placeins}
\usepackage{float}
\usepackage{booktabs}

\usepackage{xspace}
\usepackage{hyperref}
\usepackage[nameinlink]{cleveref}
\usepackage{bookmark}
\usepackage{units}
\usepackage{xcolor}

\usepackage{multirow}

\newcolumntype{C}{>{$}c<{$}} \AtBeginDocument{ \heavyrulewidth=.08em \lightrulewidth=.05em \cmidrulewidth=.03em \belowrulesep=.65ex \belowbottomsep=0pt \aboverulesep=.4ex \abovetopsep=0pt \cmidrulesep=\doublerulesep \cmidrulekern=.5em \defaultaddspace=.5em }

\graphicspath{{./figures/}}

\usepackage{xifthen}
\usepackage{xcolor}

\def\Eq#1{Eq.~\labelcref{#1}} 

 \def\Fig#1{Fig.~\labelcref{#1}}  \def\Figs#1#2{Figs.~\labelcref{#1} and \labelcref{#2}}   
\def\App#1{Sec.~\labelcref{#1}}

\newcommand{\gettitle}{Generative Modeling of Stochastic Dynamics for Long-Time Evolution}

\hypersetup{ colorlinks, linkcolor={blue!75!black}, citecolor={blue!75!black}, urlcolor={blue!75!black},
pdftitle={\gettitle}, pdfauthor={}, pdfkeywords={}, bookmarksopen=true, bookmarksopenlevel=2, bookmarksnumbered=true }

\begin{document}
\preprint{RIKEN-iTHEMS-Report-26}

\title{\gettitle}

\author{Yang-yang Tan}
\email{yangyang-tan@foxmail.com}
\affiliation{Institute for Physics of Intelligence, Graduate School of Science, The University of Tokyo, Bunkyo-ku, Tokyo 113-0033, Japan}
\affiliation{RIKEN Center for Interdisciplinary Theoretical and Mathematical Sciences (iTHEMS), Wako, Saitama 351-0198, Japan}

\author{Jinyang Li}
\email{jinyang.li@riken.jp}
\affiliation{RIKEN Center for Interdisciplinary Theoretical and Mathematical Sciences (iTHEMS), Wako, Saitama 351-0198, Japan}
\affiliation{KEK Theory Center, Institute of Particle and Nuclear Studies}
\affiliation{Graduate University for Advanced Studies (SOKENDAI), Oho 1-1, Tsukuba, Ibaraki 305-0801, Japan}

\author{Lingxiao Wang}
\email{lingxiao.wang@riken.jp}
\affiliation{RIKEN Center for Interdisciplinary Theoretical and Mathematical Sciences (iTHEMS), Wako, Saitama 351-0198, Japan}
\affiliation{Institute for Physics of Intelligence, Graduate School of Science, The University of Tokyo, Bunkyo-ku, Tokyo 113-0033, Japan}

\date{\today}

\begin{abstract}
  Exact stochastic equations for non-equilibrium dynamics are rarely accessible. We show that the long-time evolution of stochastic dynamics can be predicted from configuration pairs at a fixed short time lag, without knowledge of the equation of motion. Generative diffusion models learn the finite-time transition kernel from these pairs, and iterating it propagates the dynamics far beyond the training lag. For two-dimensional Model B, the diffusive dynamics of a conserved order parameter, the learned kernels reproduce dynamic critical scaling and self-similar $t^{1/3}$ coarsening. Agreement with direct simulations persists on lattices twice the largest training size and for initial ensembles absent from training. For driven colloids in a periodic optical potential, ten minutes of measured trajectories suffice to predict the particle current and mean passage time over the next twenty minutes within experimental uncertainty. Short-time observations thus contain the information needed to predict emergent non-equilibrium dynamics at much longer times.
\end{abstract}

\maketitle
\emph{Introduction.}---The dynamical properties of a system are encoded in the time evolution of fluctuations, transport, and emergent structures. Experiments and microscopic simulations provide time-resolved configurations, while the effective equations governing their evolution are often unknown. From a data-driven perspective, deep learning has produced emulators for weather forecasting~\cite{Lam2023GraphCast}, fluid dynamics~\cite{Kochkov2021MLCFD}, and molecular dynamics~\cite{Klein2023Timewarp,Bigi2026Action,Shoghi2026STARMD}. Deep generative models can also sample equilibrium ensembles~\cite{Noe2019BoltzmannGenerators}, including Boltzmann distributions of lattice fields~\cite{Pawlowski:2018qxs,Albergo2019Flow,Nicoli:2020njz,Cranmer:2023xbe,Wang:2023exq,Aarts:2024rsl,Zhu:2025pmw,Aarts:2026zzr}. These approaches largely target deterministic evolution or static ensembles. In a fluctuating system, thermal noise is part of the dynamical law and shapes both relaxation and observable correlations. For systems with a few degrees of freedom, generative models can learn stochastic evolution from paired observations~\cite{Chen2024FlowMap,Liu2025ConditionalDiffusion}. Langevin inference instead assumes a Langevin equation and estimates its drift and diffusion from sampled trajectories~\cite{Frishman2020SFI,Bruckner2020ULI}. For an interacting field with many degrees of freedom, it remains open whether the dynamics and its emergent behavior can be reconstructed without assuming the form of the equation of motion.

\begin{figure}[t]
  \centering
  \includegraphics[width=\columnwidth]{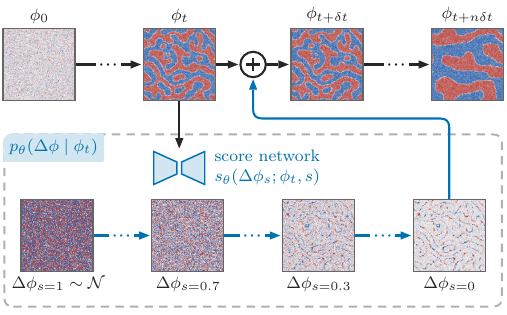}
  \caption{Conditional transition kernel. The upper row shows the physical trajectory from $\phi_0$ to $\phi_t$ and onward in steps of $\delta t$. In the lower row, the present field $\phi_t$ conditions a score network that guides reverse diffusion from Gaussian noise at $s=1$ to an increment $\Delta\phi_{s=0}\sim p_\theta(\Delta\phi\mid\phi_t)$. Adding this increment to $\phi_t$ yields $\phi_{t+\delta t}$. Repeated sampling generates $\phi_{t+n\delta t}$. The snapshots illustrate coarsening deep in the broken phase.}
  \label{fig:schematic}
\end{figure}

For autonomous dynamics, this reconstruction can be formulated in terms of a transition kernel at a fixed, finite lag $\delta t$. The same kernel applies at every $t$, and its $n$-fold composition gives the exact transition law over $n\delta t$. Here we learn this transition kernel from configuration pairs separated by $\delta t$ using a conditional diffusion model (DM)~\cite{SongErmon2019,Ho2020DDPM,Song2021SDE}. Conditioned on the present field, the model samples its stochastic increment over that interval, as illustrated in \Fig{fig:schematic}. The decisive test of a learned kernel is whether it reproduces the physics at long times and for initial ensembles and system sizes absent from training.

\begin{figure*}[t]
  \centering
  \includegraphics[width=1.0\textwidth]{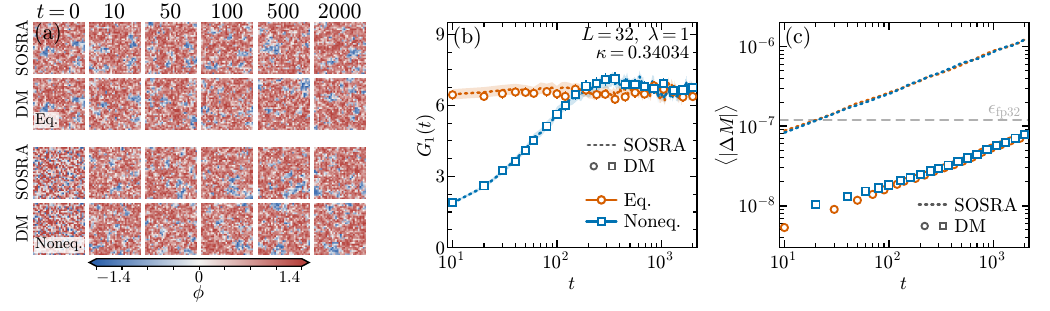}
  \caption{Long-time prediction of the learned kernel from the diffusion model (DM) 
  against a high-precision stochastic equation solver (SOSRA~\cite{Rackauckas2020SOSRA}) for Model B at the critical point
  ($\kappa=0.34034$, $\lambda=1$, $L=32$, and $\delta t=10$).
  (a)~Field configurations at $t=0$, $10$, $50$, $100$, $500$, and
  $2000$ for equilibrium HMC (upper pair) and non-equilibrium
  Gaussian (lower pair, with the spatial mean of the paired HMC field)
  initial ensembles. Within each pair, SOSRA and DM
  are shown above and below, respectively, and start from the same field at
  $t=0$.
  (b)~Equal-time power $G_1(t)$ of the smallest diagonal mode.
  Dotted lines with bands show SOSRA, and points show DM.
  (c)~Mean-field drift $\langle|M(t)-M(0)|\rangle$. Both dynamics
  conserve $M$ to single-precision accuracy. Line and marker styles are as in (b).}
  \label{fig:validation_L32}
\end{figure*}

As the first stringent test, we take two-dimensional Model B~\cite{HohenbergHalperin1977} of the Hohenberg--Halperin classification. It provides a diffusive description of critical relaxation and phase separation in binary alloys~\cite{Cahn1961} and fluid mixtures~\cite{Bray1994}, and of net-baryon fluctuations in relativistic heavy-ion collisions~\cite{Nahrgang:2018afz}. The conservation law slows the critical dynamics to $z=4-\eta$~\cite{Halperin:1974zz,Halperin:1976zz} and concentrates the difficulty in the longest wavelengths. It also freezes the spatial mean, $M(t)=M(0)$, for every realization of the noise. In training, the generative model receives only the paired fields, no prior knowledge of the equation of motion or the equilibrium weight.

The learned kernels predict the dynamics over times up to $200\,\delta t$, with each kernel trained at a single lag $\delta t$. The critical scaling analysis gives an effective dynamic exponent $z=3.792(19)$, against $3.812(20)$ from direct integration. Relaxation times also agree with direct integration on a larger lattice excluded from training. Deep in the broken phase, a kernel trained on lattices up to $L=64$ generates phase separation on $L=128$ from a disordered start. The domain size follows the $t^{1/3}$ growth law of phase ordering~\cite{Bray1994} and stays within about two percent of the direct dynamics. These scaling laws are consequences of the equation of motion, and none of them can be read off the training pairs directly. Any autonomous stochastic dynamics that can be sampled in pairs, in an experiment or in a microscopic simulation, defines the same learning problem. A system with an explicit drive or with correlated noise can often be brought into this class by enlarging the set of dynamical variables. Applied to measured trajectories of driven colloids in a periodic optical potential~\cite{dieball2025precisely}, the same construction predicts the particle current and mean passage time within experimental uncertainty (\emph{End Matter}).

\emph{Transition kernel.}---Consider a field $\phi(t)$ evolving under an autonomous stochastic equation of motion
\begin{equation}
 \partial_t\phi(t) = \mathcal F[\phi(t),\xi(t)], \label{eq:generic_dynamics}
\end{equation}
where $\xi(t)$ denotes the stochastic noise.  Neither $\mathcal F$ nor the probability law of $\xi$ depends on the choice of time origin. Equivalently, the noise path measure is invariant under a common translation of all time arguments. For a general time interval $\Delta t$ and an initial configuration $\phi(t)=\phi$, let $w$ denote a realization of the stochastic driving over $[t,t+\Delta t]$. For fixed $w$, the solution of \Eq{eq:generic_dynamics} defines the pathwise increment
\begin{equation}
 \Delta\phi(t;w)
 \equiv \phi_w(t+\Delta t)-\phi
 =\mathcal T_{\Delta t}[\phi,w],
 \label{eq:random_map}
\end{equation}
where $\phi_w(\tau)$ is the trajectory generated by the driving history $w$. Averaging over these histories gives the finite-time transition kernel
\begin{equation}
 p_{\Delta t}(\phi'\mid\phi)
 =
 \left\langle
 \delta\!\left[
 \phi'-\phi-\mathcal T_{\Delta t}[\phi,w]
 \right]
 \right\rangle_w .
 \label{eq:transition}
\end{equation}
Thus $\mathcal T_{\Delta t}$ specifies the update for a fixed noise history, and $p_{\Delta t}$ describes the distribution of possible updates. We learn $p_{\delta t}$ from paired observations at the fixed training lag $\delta t$.

For simplicity, we begin with temporally white noise. The driving histories $w$ in \Eq{eq:random_map} are then independent over successive intervals, so $\phi$ is a Markov state. Together with autonomy, this makes the transition law depend only on the interval, not on the origin. Composing the exact kernel therefore propagates the dynamics over arbitrarily long times, whereas a learned kernel must be tested for accumulated errors. Neither $\mathcal F$ nor $p_{\Delta t}$ depends on the initial ensemble. The distinction between equilibrium fluctuations and non-equilibrium relaxation therefore enters through the distribution of initial states, rather than through the transition kernel itself. The construction is not restricted to white noise. Colored noise can be represented by additional dynamical variables driven by white noise, which enlarges the state but leaves the method unchanged. A pathwise derivation and the extension to colored noise are given in \App{app:kernel_composition}.

\emph{Conditional diffusion model.}---Training data consist of pairs $(\phi_t,\phi_{t+\delta t})$ sampled from the underlying dynamics at a fixed lag $\delta t$. Taking the present configuration $\phi\equiv\phi_t$ as the condition, the model learns the conditional distribution of the increment $\Delta\phi\equiv\phi_{t+\delta t}-\phi_t$,
\begin{equation}
 p_\theta(\Delta\phi\mid\phi)\simeq
 p_{\delta t}(\phi+\Delta\phi\mid\phi).
 \label{eq:learned_transition}
\end{equation}
Unlike standard regression, which predicts only the conditional mean, $p_\theta$ represents the fluctuations that determine equal-time and temporal correlations. We use a score-based diffusion model trained on $\Delta\phi$ and conditioned on $\phi$. Repeated draws at fixed $\phi$ produce distinct increments corresponding to different stochastic realizations. The forward diffusion, score-matching objective, and reverse sampling construction are given in \App{app:conditional_construction}.

During training, the model receives only paired fields, without access to the underlying drift, noise statistics, or equilibrium weight. The input therefore has the same form as paired configurations obtained from experiments or molecular dynamics simulations. The training conditions and their coverage of field space are detailed in \App{app:numerical_details}.

A sampled increment advances the field by the training lag $\delta t$,
\begin{equation}
 \phi_{(n+1)\delta t}=\phi_{n\delta t}+\Delta\phi_n,
 \qquad
 \Delta\phi_n\sim p_\theta(\,\cdot\mid\phi_{n\delta t}).
 \label{eq:trajectory}
\end{equation}
Iterating \Eq{eq:trajectory} composes the learned kernel by the Chapman--Kolmogorov composition and generates trajectories far beyond the training lag, as illustrated in \Fig{fig:schematic}. We next test this construction on models of dynamic critical behavior where universal scaling provides a stringent long-time benchmark.

\emph{Models of dynamic critical behavior.}---Following Hohenberg and Halperin~\cite{HohenbergHalperin1977}, dynamic critical behavior is governed by the relevant slow fields $\phi=\{\phi_a(x,t)\}$. These include the order parameter, conserved densities, and hydrodynamic variables associated with broken continuous symmetries. Their long-wavelength components relax slowly, while the remaining microscopic degrees of freedom equilibrate rapidly and enter through kinetic coefficients and stochastic forces. The coarse-grained evolution takes the general Langevin form
\begin{align}
 \partial_t\phi_a(x,t)
 &=-\sum_b\mathcal K_{ab}[\phi]\, \frac{\delta F[\phi]}{\delta\phi_b(x,t)} +V_a[\phi]+\zeta_a ,
 \label{eq:general_langevin}
\end{align}
where $\zeta_b'\equiv\zeta_b(x',t')$ and $F[\phi]$ is the static free-energy functional. $ \big\langle\zeta_a\,\zeta_b'\big\rangle =2T\mathcal K_{ab}\, \delta^{(d)}(x-x')\,\delta(t-t')$ gives the \emph{fluctuation--dissipation relation}.
The kinetic operator $\mathcal K_{ab}$ governs dissipative relaxation and transport.
In the simplest diagonal case with constant $\Gamma_a>0$, $\mathcal K_{ab}=\Gamma_a\delta_{ab}$ describes local relaxation of nonconserved fields, while $\mathcal K_{ab}=-\Gamma_a\delta_{ab}\nabla^2$ describes conserved transport.
For conserved fields, the reversible term and noise must also take the form of current divergences.

The reversible term $V_a$ contains mode couplings generated by the Poisson brackets
$Q_{ab}(x,x')=\{\phi_a(x),\phi_b(x')\}$ of the slow variables,
\begin{equation}
 V_a(x)=-\sum_b\int d^dx'\left[
 Q_{ab}\,\frac{\delta F}{\delta\phi_b(x')}
 -T\,\frac{\delta Q_{ab}}{\delta\phi_b(x')}\right],
 \label{eq:mode_coupling}
\end{equation}
and does not contribute to entropy production. Equations
\eqref{eq:general_langevin} and \eqref{eq:mode_coupling} make
$P_{\rm eq}[\phi]\propto e^{-F[\phi]/T}$ stationary. Thus $F$
determines the statics, whereas the dynamics also depends on the slow fields, the kinetic operator $\mathcal K_{ab}$, and the mode couplings $Q_{ab}$.

These ingredients define the dynamic universality classes. Models A and B~\cite{Halperin:1976zz} contain only a nonconserved or conserved order parameter, respectively, with $V_a=0$. Models C and D~\cite{Halperin:1974zz} couple the order parameter dissipatively to a conserved scalar density. Models E, F, G, H, and J~\cite{Halperin:1976zza,Siggia:1976zz,MaMazenko1975} have $Q_{ab}\neq0$, corresponding to reversible couplings involving conserved densities, symmetry generators, or transverse momentum. Systems in the same static universality class may therefore relax with different dynamic critical exponents, $t_{\mathrm{relax}}\sim\xi^z$. For autonomous Markovian models, the kernel is conditioned on the complete set of slow fields.

\begin{figure}[t]
  \centering
  \includegraphics[width=\columnwidth]{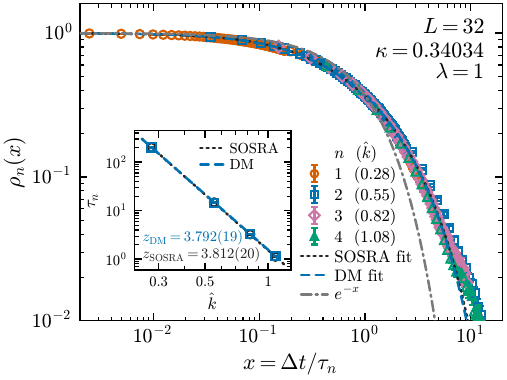}
  \caption{Critical relaxation at $\kappa=0.34034$, $\lambda=1$, and $L=32$. The diagonal-mode correlations collapse when plotted against $x=\Delta t/\tau_n$. Colored markers show the DM results, with error bars denoting standard errors. The dotted black curve and the dashed blue curve show fits of \Eq{eq:critical_fit_form} to the SOSRA and DM data, respectively. The gray dash-dotted curve is $e^{-x}$. The inset shows the relaxation times versus lattice momentum, yielding effective exponents $z=3.792(19)$ for the DM and $z=3.812(20)$ for SOSRA.}
  \label{fig:critical_scaling}
\end{figure}

For scalar Model B, we take $\mathcal K=-\Gamma\nabla^2$, $V=0$, and a continuum $\phi^4$ free energy $F[\phi]$. For periodic or no-flux boundaries, the mean field
\begin{equation}
 M(t)=\frac{1}{|\Omega|}\int_\Omega d^dx\,\phi(x,t)=M(0)
 \label{eq:modelb_conservation}
\end{equation} is fixed for every noise realization, while the nonzero modes relax. We enforce conservation by projecting each generated increment onto the zero-mean subspace, preserving $M$ to floating-point accuracy. The lattice realization and implementation details are given in \App{app:numerical_details}. 

Trained on pairs generated from synthetic correlated Gaussian fields, the kernel reproduces long-time evolution from equilibrium and non-equilibrium initial ensembles absent from training. \Fig{fig:validation_L32} compares its predictions at $L=32$ with direct
integration using the stability-optimized stochastic Runge--Kutta scheme
SOSRA~\cite{Rackauckas2020SOSRA}. The comparison includes an equilibrium
hybrid Monte Carlo (HMC) ensemble~\cite{Duane:1987de} and an uncorrelated
Gaussian ensemble far from equilibrium. For each ensemble, the learned
and reference dynamics started from the same fields and were evolved to
$t=2000$. For the diagonal modes $k_n=(2\pi n/L)(1,1)$, the equal-time
power is $G_n(t)=\langle|\phi(k_n,t)|^2\rangle/V$, where
$\phi(k,t)=\sum_x e^{-ik\cdot x}\phi(x,t)$ and $V=L^2$.
The generated configurations reproduced the large-scale
structure of the reference dynamics, while $G_1(t)$ tracked the direct
result both in equilibrium and during
relaxation from the Gaussian ensemble. The mean field remained conserved
to single-precision accuracy. Temporal correlations of the first four
diagonal modes agreed with direct integration over the resolved range of
time separations
(\Fig{fig:rho_L32}). Details of the predictions are given in
\App{app:observable_details}.

\emph{Critical relaxation and coarsening.}---We characterize relaxation by the normalized
correlation
\begin{equation}
 \rho_n(\Delta t)=
 \frac{V^{-1}\left\langle\operatorname{Re}\!\left[
 \phi(k_n,t+\Delta t)\,\phi(k_n,t)^*
 \right]\right\rangle_t}
 {\sqrt{\left\langle G_n(t+\Delta t)\right\rangle_t
 \left\langle G_n(t)\right\rangle_t}},
 \label{eq:rho_def}
\end{equation}
where $\Delta t$ is the time separation and $\langle\cdot\rangle_t$ averages
over trajectories and all time origins with that separation. On trajectories
generated with a fixed lag $\delta t$, the available separations are
$\Delta t=m\delta t$ with $m=0,1,2,\ldots$. We define $\tau_n$ by
$\rho_n(\tau_n)=e^{-1}$. At the critical point, dynamic scaling predicts
\begin{equation}
 \rho_n(\Delta t)=f(\Delta t/\tau_n),
 \qquad
 \tau_n\propto \hat k_n^{-z},
 \label{eq:rho_scaling}
\end{equation}
where $\hat k_n$ is the lattice momentum magnitude at $k_n$, with
$\hat k^2=\sum_{\mu=1}^{2}\hat k_\mu^2$ and
$\hat k_\mu=2\sin(k_\mu/2)$.

The critical scaling analysis in \Fig{fig:critical_scaling} uses kernels
trained on equilibrium pairs at $L=8,16,32$, with training lags
$\delta t=0.5$ and $10$ to resolve the fast and slow modes. The learned and direct
correlations approximately collapse onto the same nonexponential profile,
with slower decay than $e^{-x}$ at large $x=\Delta t/\tau_n$. We describe
this profile by
\begin{equation}
 f(x)\simeq e^{-\lambda_0 x}(1+ax)^{-b}.
 \label{eq:critical_fit_form}
\end{equation}
Its effective decay rate $-d\ln f/dx$ decreases from $\lambda_0+ab$
toward $\lambda_0$. We constrain the initial rate using the static
fluctuations and the equilibrium fluctuation-dissipation relation
(\App{app:dynamic_scaling_function}). Reproducing this crossover tests
the learned dynamics beyond the critical exponent.

Fits to the relaxation times yield effective exponents $z=3.792(19)$ for the learned kernels
and $z=3.812(20)$ for SOSRA, consistent within the quoted uncertainties.
These values characterize the resolved momentum range; the asymptotic
two-dimensional value is $z=4-\eta=15/4$ (\App{app:multiL}).

The same equilibrium-trained kernels were then applied at the larger size $L=64$. Across the resolved momentum range, the learned relaxation times agreed with the direct results, yielding $z=3.855(21)$ for the learned kernel and $3.867(23)$ for the direct dynamics (\Fig{fig:reverse_nfe_tau_L64}).
\begin{figure}[t]
 \centering
 \includegraphics[width=\columnwidth]{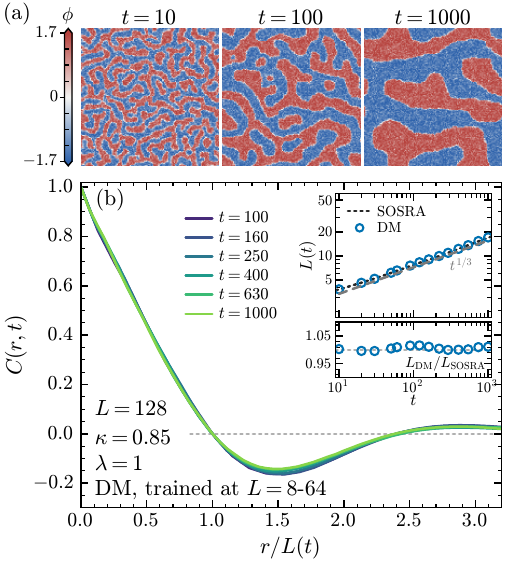}
 \caption{Coarsening deep in the broken phase
 ($\kappa=0.85$, $\lambda=1$). The kernel was trained only on synthetic
 field configurations at $L\leq64$ and applied at $L=128$ from a
 disordered initial state. (a)~DM configurations at $t=10$, $100$, and
 $1000$. (b)~Collapse of the sign-field correlation $C(r,t)$ for
 $t=100$--$1000$. The insets compare the domain size $L(t)$, defined by
 the first zero of $C(r,t)$, with the direct SOSRA result and show their
 ratio. The black dotted line shows SOSRA, and blue open circles
 show DM. The gray dashed line denotes the $t^{1/3}$ growth law. The SOSRA
 and DM ensembles contain $128$ and $32$ trajectories, respectively.}
 \label{fig:collapse_L128}
\end{figure}

Deep in the broken phase, a separate kernel trained on synthetic domain
fields at $L\leq64$ was applied at $L=128$ from a disordered initial
state. The resulting fields underwent
phase separation. Domain growth was quantified by the equal-time
correlation of the sign field
$\sigma(x,t)=\operatorname{sgn}\phi(x,t)$~\cite{PuriBrayLebowitz1997},
\begin{equation}
 C(r,t)=\left\langle\sigma(x,t)\,\sigma(x+r,t)\right\rangle,
 \label{eq:sign_correlation}
\end{equation}
where the average is over lattice sites, orientations of $r$, and
trajectories. Because $\sigma=\pm1$, $C(0,t)=1$ without further
normalization. The domain size $L(t)$ is defined as the first zero of
$C(r,t)$~\cite{Zhu1999}. For a conserved scalar order parameter, phase-ordering scaling
predicts~\cite{LifshitzSlyozov1961,Huse1986,Rogers1988,Bray1994}
\begin{equation}
 C(r,t)=g\big(r/L(t)\big),
 \qquad
 L(t)\propto t^{1/3}.
 \label{eq:coarsening_scaling}
\end{equation}

Between $t=100$ and $1000$, correlations from the learned dynamics collapsed onto a single scaling curve (\Fig{fig:collapse_L128}). The corresponding domain size followed the $t^{1/3}$ growth law and remained within about two percent of the direct Model B result. The structure factor also exhibited the Porod scaling expected for sharp domain walls~\cite{Porod1951,Bray1994} (\App{app:observable_details}).

\emph{Discussion.}---The dynamic exponent and coarsening law are collective consequences of the dynamics, but neither is imposed during training. Each kernel is learned from configuration pairs at a fixed lag, of the kind obtainable from experiments or microscopic simulations, without long-time trajectories or scaling observables as training targets. Recovering these laws through repeated application of the learned kernels provides a nontrivial test of their ability to reproduce long-time collective dynamics.

The same construction can be used whenever the measured variables form a sufficient state. A reconstruction from trajectories of driven colloids in a periodic optical potential is reported in the End Matter~\cite{dieball2025precisely}. Beyond the training window, the learned kernel reproduces the measured particle current and mean passage time under driving to within $1\%$, whereas a Langevin reference based on Ref.~\cite{dieball2025precisely} deviates by about $6\%$.

The same construction can be extended to systems with memory by including
suitable auxiliary variables in the state~\cite{Mori:1965oqj}.
Active and reaction--diffusion systems may require additional slow fields
in the conditioned state~\cite{TonerTu1995,Turing1952}.
Periodic drives can be included through the drive phase~\cite{FroylandKoltai2017}.
The construction may also be extended to other Hohenberg--Halperin classes.
Model A describes local relaxation of a
nonconserved order parameter~\cite{Halperin:1976zz,Tan:2025bsv,Tan:2024fuq}.
Model H couples a conserved order parameter to conserved momentum through
reversible terms~\cite{Chen:2024lzz,Roth:2024hcu,Chattopadhyay:2024jlh}.
The resulting interplay of diffusion and advection governs
critical relaxation in fluids and at the
critical end point of quantum chromodynamics~\cite{HohenbergHalperin1977,Son:2004iv}.
Beyond critical phenomena, the same approach could be applied to stochastic
hydrodynamics and molecular dynamics.

\emph{Conclusions.}---Finite-time transition kernels can be learned from paired observations and iterated without an explicit equation of motion. In Model B, this reconstruction recovered critical relaxation, transfer to a larger lattice, and $t^{1/3}$ phase ordering far beyond the training lag. Trained on experimental trajectories of driven colloids, it predicted the particle current and mean passage time within experimental uncertainty, more accurately than a Langevin reference. How far this extends depends on which slow variables a pair of observations has to carry, and that choice differs for every dynamic universality class.

\emph{Acknowledgments.---} We thank Gert Aarts, Yuto Ashida, Kenji Fukushima, Satoshi Iso, Jan M. Pawlowski and other DM-QFT collaboration members for insightful discussions.
We thank the DEEP-IN working group at RIKEN-iTHEMS for support in the preparation of this paper. This research is partially supported by Initiative on Recommendation Program for Young Researchers and Woman Researchers, Information Technology Center, The University of Tokyo. We also thank the support from Google DeepMind team.
LW and YT are supported by the JST-BOOST Grant No.\ JPMJBY24H9. LW is also supported by the RIKEN-TRIP initiative (RIKEN-Quantum) and JSPS KAKENHI Grant No.\ 25H01560.

\emph{Code availability.---}The code used in this work will be made publicly available upon publication.

\emph{Use of AI tools.---}GPT 6 Astra (OpenAI) and Claude Fable 5 (Anthropic) assisted substantially with numerical cross-checks and with polishing the manuscript. All derivations, results, and statements were verified by authors, who take full responsibility for the content.

\bibliography{references}

\clearpage
\begingroup
\setlength{\textfloatsep}{6pt}

\section*{End Matter}

\emph{Computational cost.} We benchmark the learned kernel against leading
publicly available stochastic solvers on the same NVIDIA GH200 GPU
(Table~\ref{tab:gpu_wall_time}). For the same ensembles and evolution time at
comparable accuracy, the learned kernel is faster than every public
implementation tested. Our fused SOSRA implementation is faster in
equilibrium and comparable in cost out of equilibrium.

\emph{Colloid experiment.} We apply our framework to an experiment on colloidal particles in a periodic potential~\cite{dieball2025precisely,Dieball2024data}. The particles diffuse in a quasi-two-dimensional cell
subject to a periodic optical potential along $x$. In the absence of
external driving, thermal fluctuations produce motion within individual
wells together with activated hopping between neighboring wells.
Translating the sample cell relative to the optical pattern biases these
hopping processes and generates a non-equilibrium particle current. The
same system therefore probes both thermally activated transport and
driven stochastic dynamics.

\onecolumngrid
\noindent\includegraphics[width=\textwidth]{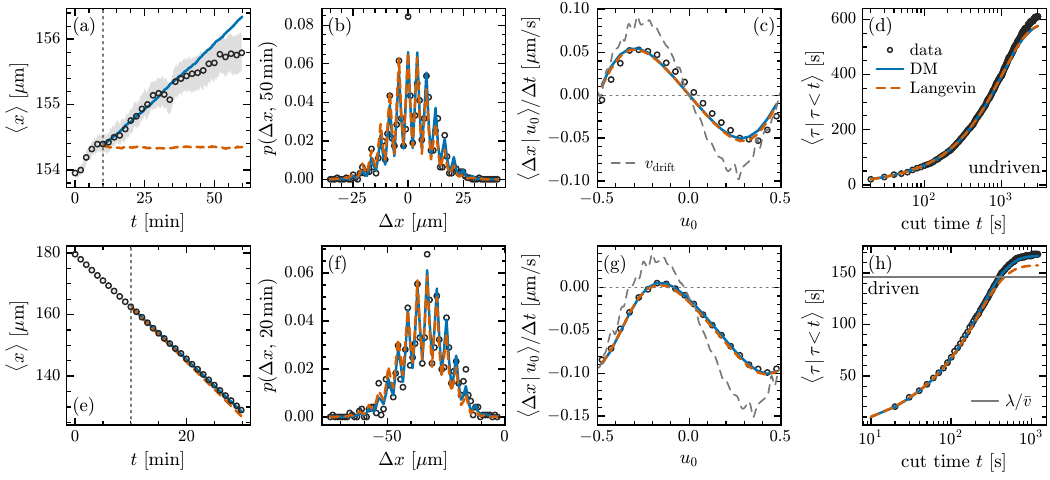}\par
\refstepcounter{figure}\label{fig:experiment}
\makeatletter\@makecaption{\fnum@figure}{DM predictions for undriven
 (upper row) and driven (lower row) colloidal motion in a periodic
 potential~\cite{dieball2025precisely,Dieball2024data}.
 Black circles show the experimental data. Blue solid lines show
 DM and orange dashed lines the Langevin reference based on
 Ref.~\cite{dieball2025precisely}.
 (a,e)~Mean position $\langle x(t)\rangle$. Shading shows the standard error
 of the mean displacement from $t=0$. The dotted line at $t=10$ min marks
 the end of training.
 (b,f)~Distributions of the net displacement over $50$ min of undriven
 motion and $20$ min of driven motion, both measured from $t=10$ min.
 (c,g)~Mean displacement rate $\langle\Delta x\mid u_0\rangle/\Delta t$
 at $\Delta t=\delta t=10$ s as a function of the initial phase $u_0$.
 Gray dashed lines show the instantaneous Langevin drift $v_{\rm drift}$,
 corresponding to $\Delta t\to0$.
 (d,h)~Mean passage time $\langle\tau\mid\tau<t\rangle$ conditioned
 on passage before $t$. The target lies one potential period $\lambda$
 from the initial position, in either direction in (d) and along the
 drive in (h). Passage times are evaluated at $10$ s resolution.
 The gray line in (h) shows the continuous-time mean passage time,
 the period divided by the mean speed during training.}\makeatother
\twocolumngrid

Separate kernels are learned for the undriven and driven dynamics from
position pairs separated by $\delta t=10$ s, conditioned on the particle’s
phase within one period of the optical potential. Starting from the
measured positions at $t=10$ min, the kernels are iterated to predict the
subsequent $50$ min of undriven motion and $20$ min of driven motion.

DM retains the weak net drift in the undriven experiment
[\Fig{fig:experiment}(a)], which the equilibrium Langevin reference based on Ref.~\cite{dieball2025precisely} omits.
This effect is learned from the measured motion without specifying its
physical origin. Under driving, the predicted mean speed agrees with
experiment to within $1\%$ [\Fig{fig:experiment}(e)].

DM also closely reproduces the measured displacement distributions
[\Fig{fig:experiment}(b,f)] and passage-time statistics
[\Fig{fig:experiment}(d,h)]. Its overall accuracy is comparable to that of
the Langevin reference, with closer agreement for the driven mean current and mean passage time
[\Fig{fig:experiment}(e,h)]. Further comparisons are given in \App{app:experiment}.

\begin{table}[H]
\caption{Wall time for evolving 512 configurations on an $L=64$ lattice from $t=0$ to $100$, for the equilibrium (Eq.) HMC ensemble and the non-equilibrium (Noneq.) quench ensemble of \Fig{fig:collapse_L128}. Details are given in \App{app:timing_benchmark} and \App{app:non-equilibrium_timing}.}
\label{tab:gpu_wall_time}
\begin{ruledtabular}
\begin{tabular}{lcc}
& \multicolumn{2}{c}{Wall time (s)} \\
\cline{2-3}
Method & Eq. & Noneq. \\
\hline
DPM-Solver-v3 (This work) & $3.24$ & $3.26$ \\
SOSRA (This work) & $1.30$ & $3.28$ \\
SOSRA (StochasticDiffEq.jl)~\cite{Rackauckas2020SOSRA,Rackauckas2017DifferentialEquations} & $12.36$ & $30.67$ \\
ShARK (Diffrax)~\cite{Foster2024ShARK,Kidger2021Diffrax} & $5.56$ & $17.76$ \\
EM (Diffrax)~\cite{Maruyama1955,Kidger2021Diffrax} & $19.45$ & $38.75$ \\
EM (other GPU/CPU solvers)~\cite{Maruyama1955} & $>100$ & $>200$ \\
\end{tabular}
\end{ruledtabular}
\end{table}

\clearpage
\endgroup
\onecolumngrid
\begin{center}
 \textbf{\large Supplemental Material for ``\gettitle''}
\end{center}
\setcounter{equation}{0}
\setcounter{figure}{0}
\setcounter{table}{0}
\setcounter{section}{0}
\renewcommand{\theequation}{S\arabic{equation}}
\renewcommand{\thefigure}{S\arabic{figure}}
\renewcommand{\thetable}{S\arabic{table}}
\renewcommand{\thesection}{S\arabic{section}}
\renewcommand{\theHequation}{S\arabic{equation}}
\renewcommand{\theHfigure}{S\arabic{figure}}
\renewcommand{\theHtable}{S\arabic{table}}
\renewcommand{\theHsection}{S\arabic{section}}
\setcounter{secnumdepth}{2}
\twocolumngrid

\section{Finite-time composition and colored noise}
\label{app:kernel_composition}

The finite-time construction requires a state that determines the
conditional law of future configurations. For the white-noise dynamics
studied here, this state is $\phi$.

Let $w_1$ and $w_2$ denote the driving histories over successive intervals
of lengths $\Delta t_1$ and $\Delta t_2$, expressed relative to the start of
each interval. Their concatenation is $w_1\oplus w_2$. With the intermediate
state $\phi'=\phi+\mathcal T_{\Delta t_1}[\phi,w_1]$, causality and uniqueness
give
\begin{equation}
 \begin{aligned}
 \mathcal T_{\Delta t_1+\Delta t_2}[\phi,w_1\oplus w_2]
 &=\mathcal T_{\Delta t_1}[\phi,w_1]
 \\
 &\quad+\mathcal T_{\Delta t_2}[\phi',w_2].
 \end{aligned}
 \label{eq:pathwise_composition}
\end{equation}
This identity holds for each realization before averaging over the
driving. For a complete Markov state, the conditional law of the second
update depends on the past only through $\phi'$. Autonomy makes that law
independent of the time origin. Averaging the two successive updates
therefore gives the Chapman--Kolmogorov identity
\begin{equation}
 p_{\Delta t_1+\Delta t_2}(\phi''\mid\phi)
 =\int\!\mathcal D\phi'\,
 p_{\Delta t_2}(\phi''\mid\phi')p_{\Delta t_1}(\phi'\mid\phi).
 \label{eq:kernel_composition}
\end{equation}
Independent draws from the exact kernel $p_{\delta t}$, conditioned on the current
state, thus reproduce the joint distribution of configurations at
$0,\delta t,2\delta t,\ldots$ for any chosen initial ensemble. This result
concerns the sampled times and does not specify the intervening
trajectories. For a learned kernel, small errors in each update can
accumulate over successive intervals. Agreement over one interval
therefore does not guarantee accurate long-time dynamics. This must be
tested by comparing observables from generated trajectories with
measurements or direct simulations.

For colored noise, let $\eta$ denote the dynamical variables that carry
its temporal correlations and use the joint state $X=(\phi,\eta)$.
When this state is autonomous and Markovian, its transition kernel
$p^X_{\Delta t}(X'\mid X)$ obeys \Eq{eq:kernel_composition} with $\phi$
replaced by $X$. The learning construction is unchanged, with training
pairs $(X_t,X_{t+\delta t})$. Each sampled update advances both the field
and the noise variables, preserving the memory between successive
physical intervals.

\section{Conditional diffusion construction}
\label{app:conditional_construction}

\begin{figure*}[t]
 \centering
 \includegraphics[width=\textwidth]{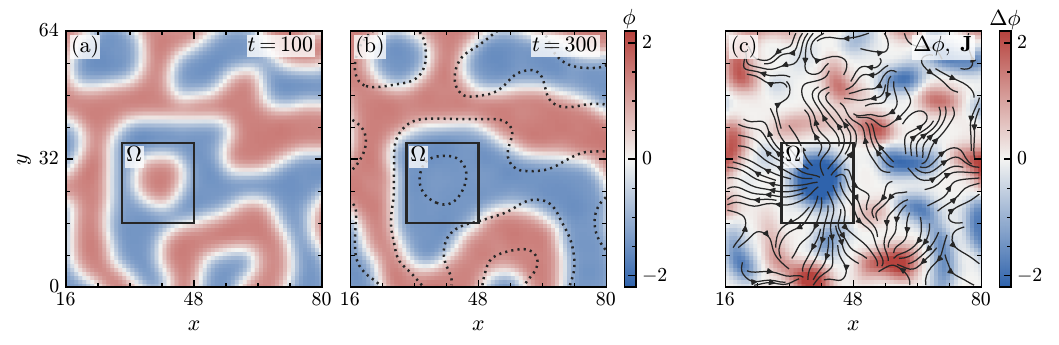}
 \caption{Conserved transport during phase ordering on an $L=128$ lattice
 at $\kappa=0.85$ and $\lambda=1$. All panels show the same
 enlarged region of one SOSRA trajectory. (a,b)~Field $\phi$ at $t=100$
 and $300$. The dotted line in (b) marks $\phi(t=100)=0$.
 (c)~Colors show the field change $\Delta\phi=\phi(300)-\phi(100)$.
 Selected streamlines show the direction of the integrated current
 $\mathbf J=\int_{100}^{300}\mathbf j\,dt$, including the noise contribution
 in \Eq{eq:conserved_current}.
 Very weak currents are omitted from the streamline overlay.
 The fixed box $\Omega$ encloses a shrinking
 domain. Fields and current are smoothed with the same periodic Gaussian
 of standard deviation two lattice spacings, preserving the discrete
 continuity relation $\Delta\phi+\nabla\cdot\mathbf J=0$.}
 \label{fig:conserved_current}
\end{figure*}

For each observed pair, the present state $\phi$ is the condition and
$\Delta\phi$ is the target increment. Predicting an increment is convenient
for slowly evolving fields, although the construction also applies to the
next state itself. For coupled fields, $\phi=\{\phi_a\}$ contains all
variables required for a Markov description.

To represent $p_\theta$ we diffuse only the increment $\Delta\phi$ and pass the
condition $\phi$ to the score network as an unperturbed input.  A
variance-exploding (VE) process~\cite{Song2021SDE},
\begin{equation}
 d\Delta\phi_s=g(s)\,dW_s,
 \qquad
 \Delta\phi_s=\Delta\phi+\Sigma(s)\eta,
 \label{eq:ve_kernel}
\end{equation}
with auxiliary diffusion time $s\in[0,1]$, noise
$\eta\sim\mathcal N(0,\mathbb 1)$, and
$\Sigma^2(s)=\int_0^s g^2(s')\,ds'$, defines a family of noisy conditional
distributions.
The auxiliary time $s$ is distinct from the physical time $t$ and the
fixed training lag $\delta t$. We train the score by denoising score
matching~\cite{Vincent2011},
\begin{equation}
 \mathcal L(\theta)=
 \big\langle
 \left\|\Sigma(s)\,s_\theta
 (\Delta\phi+\Sigma(s)\eta,\phi,s)+\eta\right\|^2
 \big\rangle_{s,\phi,\Delta\phi,\eta}.
 \label{eq:dsm_loss}
\end{equation}
The average is over the measured pairs, the auxiliary time, and the added
Gaussian noise. The objective contains neither the physical drift nor the
physical noise covariance. The physical dynamics enters through the pairs.

One finite physical-time increment is drawn by integrating the conditional
reverse-time SDE~\cite{Anderson1982,Song2021SDE},
\begin{equation}
 d\Delta\phi_s=-g(s)^2 s_\theta(\Delta\phi_s,\phi,s)\,ds
 +g(s)\,dW_s .
 \label{eq:reverse_sde}
\end{equation}
The equation is integrated from $s=1$ toward $s=0$ at fixed $\phi$.
Here $W_s$ denotes the Wiener process in this direction, with increment
covariance $|ds|\mathbb 1$.
Sampling starts from a Gaussian whose variance is set by $\Sigma^2(1)$.
This approximates the terminal forward distribution when the added noise
dominates the target increment. The sampled increment is added to the
condition according to \Eq{eq:trajectory}. At the next physical step the
condition is updated and a fresh history of the auxiliary noise is drawn.
The reverse-time evolution is itself a Langevin process on configuration
space whose drift is proportional to the learned score, in analogy with stochastic
quantization~\cite{Parisi:1980ys,Damgaard:1987rr}.  This correspondence has
been developed for lattice field theories
in~\cite{Wang:2023exq,Fukushima:2024oij,Hirono:2024zyg}.
Stochastic quantization has also been formulated as an optimal control
problem at finite auxiliary time~\cite{Wang:2026pmv}.
Renormalization group flows give a further reading of generative
flows~\cite{Ihssen:2024ihp,Bauer:2024byr,Masuki:2025pfx,Masuki:2026qfw}.

\section{Numerical details of the Model B case study}
\label{app:numerical_details}

\subsection{Lattice realization}

For the conserved scalar field considered in the main text,
\Eq{eq:general_langevin} reduces to
\begin{align}
 \partial_t\phi(x,t)
 &=\Gamma\nabla^2\frac{\delta F[\phi]}{\delta\phi(x,t)}
 +\zeta(x,t),
 \nonumber\\
 \big\langle\zeta\,\zeta'\big\rangle
 &=-2T\Gamma\nabla^2\delta^{(d)}(x-x')\,\delta(t-t').
 \label{eq:modelb_general}
\end{align}
We use the lattice $\phi^4$ action
\begin{equation}
S[\phi] = \sum_{x} \Bigl[
  -2\kappa \sum_{\mu=1}^{D} \phi_x\,\phi_{x+\hat\mu}
  + \phi_x^2 + \lambda\bigl(\phi_x^2 - 1\bigr)^2
\Bigr],
 \label{eq:lattice_action}
\end{equation}
on a periodic lattice with $D=2$ and unit spacing.
Here $\kappa$ is the hopping parameter, $\lambda$ the dimensionless
self-coupling, and $\hat\mu$ a lattice unit vector.

Writing $\zeta=-\nabla\cdot\boldsymbol\xi$ with Gaussian current noise
$\boldsymbol\xi$, we define
\begin{align}
 \mathbf j&=-\Gamma\nabla\frac{\delta F}{\delta\phi}+\boldsymbol\xi,
 \nonumber\\
 \partial_t\phi+\nabla\cdot\mathbf j&=0.
 \label{eq:conserved_current}
\end{align}
For periodic or no-flux boundaries, the net boundary flux vanishes.
Integrating the continuity equation gives $dM/dt=0$ in every realization.
We discretize the current with nearest-neighbor differences and use
$\Gamma=1/\gamma$.
We set $\lambda=T=\gamma=1$ throughout the Model B calculations.
We use HMC to sample equilibrium initial fields with Boltzmann weight $e^{-S}$.
\Fig{fig:conserved_current} illustrates this local conservation law during
phase ordering through the field change and the integrated total current.

Training targets are obtained by direct SOSRA integration with step size
$0.002$. For the Gaussian initial fields used to train the kernel in
\Fig{fig:validation_L32}, the first unit of physical time uses step size
$10^{-4}$ before returning to $0.002$. All generated pairs are used for
training.

\subsection{Training pairs}

The training data for the kernel used in \Fig{fig:validation_L32} of the
main text contain $1024000$ pairs at $L=32$, generated from initial fields
\begin{equation}
 \phi_0(x)=\chi(x)+M,
 \label{eq:training_condition}
\end{equation}
where $M\sim\mathcal U[-0.8,0.8]$. The mean-zero part is drawn from
\begin{equation}
 \chi(k)=
 \frac{\xi(k)}{\sqrt{Z\hat k^2+m^2}},
 \qquad k\ne0,
 \qquad
 \chi(0)=0,
 \label{eq:training_propagator}
\end{equation}
where $\xi(x)$ is independent unit-variance Gaussian noise,
$Z\sim\mathcal U[0.45,1.15]$, and $m^2\sim\mathcal U[0,0.8]$.
The three parameters are sampled independently. The initial field is evolved
over the training lag $\delta t=10$, and the target is the resulting increment.

The training data for the kernels used in
\Figs{fig:critical_scaling}{fig:reverse_nfe_tau_L64} are generated from
canonical ensembles at $L=8,16,32$, with $51200$ initial fields at each size.
No $L=64$ field is included in these training sets.
Separate kernels are trained at $\delta t=0.5$ and $10$.
The single-size kernel in \Figs{fig:L32_chain}{fig:one_step_comparison}
is trained on the $L=32$ canonical pairs alone.

The training data for the kernel used in \Fig{fig:collapse_L128} are
generated from centered synthetic domain fields at $L=8,16,32,64$. Let
\begin{equation}
 (\mathcal P_0 f)_x=f_x-\frac{1}{V}\sum_y f_y
 \label{eq:zero_mode_projector}
\end{equation}
denote the projector onto zero-mean fields. The initial fields are
\begin{equation}
 \phi_0(x)=\mathcal P_0\!\left\{
 A\tanh\!\left[\frac{\ell}{w}g_\ell(x)\right]+\eta\xi(x)\right\},
 \label{eq:domain_training_condition}
\end{equation}
where $g_\ell$ is a smoothed Gaussian field with correlation length $\ell$. We draw independent unit-variance
Gaussian noise $\zeta(x)$, multiply each Fourier mode $k\ne0$ by
$\exp[-\hat k^2\ell^2/4]$, and set the $k=0$ mode to zero so that the spatial
mean vanishes. The resulting real-space field is divided by its
root-mean-square value $[V^{-1}\sum_x g_\ell^2(x)]^{1/2}$, so that every
sample has unit variance. The hyperbolic tangent then hardens the level sets
of $g_\ell$ into domains of size $\ell$ with walls of width $w$.
The added noise $\xi$ is independent of $\zeta$.
We sample $\log\ell\sim\mathcal U[\log2,\log(L/3)]$,
$\log w\sim\mathcal U[\log0.7,\log3]$,
$A\sim\mathcal U[0.8,1.8]$, and $\eta\sim\mathcal U[0.05,0.35]$.
With probability $1/8$, we instead set $A=0$ and draw
$\eta\sim\mathcal U[0.05,0.2]$.
The target increment is obtained by evolving each synthetic field for the
training lag $\delta t=10$. We generate $51200$ pairs at each size.

Since the dynamics is symmetric under $\phi\to-\phi$, the transition
kernel satisfies $p(\Delta\phi\mid\phi_t)=p(-\Delta\phi\mid-\phi_t)$, and
every training pair is also used with both signs reversed.

\subsection{Score model and reverse sampling}

The network operates on rescaled fields
$\psi=\Delta\phi/A_{\rm res}$ and $\varphi=\phi_t/A_{\rm cond}$, where
$A_{\rm res}$ and $A_{\rm cond}$ are the largest absolute values of the
increment and of the initial field over the training set.
For the VE process we use
\begin{equation}
 g(s)=\sigma^s,\qquad
 \Sigma^2(s)=\frac{\sigma^{2s}-1}{2\log\sigma},
 \label{eq:ve_schedule}
\end{equation}
with $s\sim\mathcal U[10^{-5},1]$. We use $\sigma=250$ at
$\kappa=0.34034$ and $\sigma=190$ at $\kappa=0.85$.
The score is represented by the modified NCSN++ network introduced in
Ref.~\cite{Tan:2026ibs}. It uses periodic convolutions and omits the
internal residual connections of NCSN++~\cite{Song2021SDE}, while retaining
the encoder-to-decoder skip connections. Its inputs are the noisy increment,
the initial field, and the auxiliary time. Training uses Adam with learning
rate $10^{-3}$.

Sampling uses an exponential moving average of the network weights.
The reverse process starts from $\psi_1\sim\mathcal N(0,\Sigma^2(1)\mathbb 1)$ and is
integrated with 2000 Euler--Maruyama (EM) steps, logarithmically spaced in
$s$ down to $10^{-5}$.
Independent reverse noise is used for successive physical intervals.

Conservation is imposed only when drawing a transition, using the projector
$\mathcal P_0$ of \Eq{eq:zero_mode_projector}.
The initial Gaussian residual is projected with $\mathcal P_0$.  At every
reverse step, both the predicted score and the newly drawn Gaussian noise are
projected with the same operator.  The sampled process is therefore
\begin{equation}
 d\psi_s=
 -g^2(s)\mathcal P_0 s_\theta
 (\psi_s,\varphi,s)\,ds
 +g(s)\mathcal P_0 dW_s.
 \label{eq:projected_reverse_sde}
\end{equation}
The final residual is rescaled to physical units and projected once more
before it is added to the condition.
Consequently $\sum_x\Delta\phi_x=0$ for every generated transition and $M$ is
preserved up to floating-point roundoff. The projection fixes only the zero
mode and leaves the other modes to the learned dynamics.

The $L=64$ comparison with DPM-Solver-v3 is given in
\App{app:multiL}.

\section{Model B observables, critical relaxation, and coarsening}
\label{app:observable_details}

\begin{figure}[t]
 \centering
 \includegraphics[width=\columnwidth]{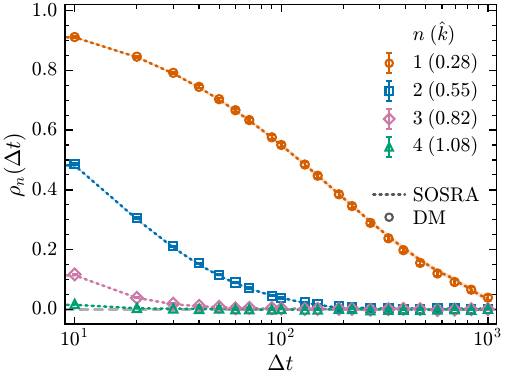}
 \caption{Time-origin averaged mode correlations $\rho_n(\Delta t)$ of the
 first four diagonal modes on the equilibrium ensemble of
 \Fig{fig:validation_L32}. Dotted lines with bands show SOSRA, and open markers
 show the DM kernel trained on Gaussian fields. All time origins from the $t=2000$
 trajectories are included. Bands and error bars are trajectory jackknife
 standard errors over $1024$ trajectories.}
 \label{fig:rho_L32}
\end{figure}

\subsection{Initial ensembles and statistical errors}

For \Fig{fig:validation_L32}, we use $1024$ equilibrium initial fields and
$1024$ uncorrelated Gaussian initial fields at $L=32$. Each Gaussian field
is shifted to have the same spatial mean as its paired equilibrium field,
so that the conserved zero mode is matched between the two ensembles. DM
and SOSRA start from the same initial configurations but use independent
stochastic noise. \Fig{fig:validation_L32} shows the evolution up to
$t=2000$. All available time origins of the equilibrium trajectories are
used to evaluate the correlations in \Fig{fig:rho_L32}.

For \Fig{fig:critical_scaling}, the two DM kernels start from the same
$2048$ equilibrium initial fields at $L=32$. The $\delta t=0.5$ kernel is
iterated to $t=50$, and the $\delta t=10$ kernel is iterated to
$t=1000$. The SOSRA reference starts from the same initial fields and is
evolved to $t=1000$ in a single run. Configurations are recorded every
$0.5$ up to $t=50$ and every $10$ thereafter, so that each learned kernel
is compared with the reference dynamics on the corresponding time grid.

For \Fig{fig:collapse_L128}, the kernel trained on synthetic domain fields
at $L\leq64$ is applied at $L=128$ and $\kappa=0.85$. We use $32$ initial
fields
\begin{equation}
 \phi_0=0.1\,\mathcal P_0\xi,
 \label{eq:quench_initial}
\end{equation}
where $\xi$ is independent unit-variance Gaussian noise. Each field is
evolved with $\delta t=10$ to $t=1000$. The SOSRA reference uses $128$
independent trajectories initialized from the same distribution.

For each mode index $n$, we average over the four symmetry-related
wave vectors of $k_n$. This average is performed
separately for the equal-time quantity $|\phi(k_n,t)|^2$ and the
unequal-time product $\operatorname{Re}[\phi(k_n,t+\Delta t)\phi(k_n,t)^*]$
before constructing $G_n$ and $\rho_n$. The bands
and error bars of $G_1(t)$ in \Fig{fig:validation_L32} show standard
errors over trajectories. For $\rho_n$ in \Fig{fig:rho_L32}, we estimate
the uncertainty with a leave-one-trajectory-out jackknife. For each
jackknife sample, the complete estimator in \Eq{eq:rho_def} is recomputed
after removing one trajectory.

\begin{figure}[t]
 \centering
 \includegraphics[width=\columnwidth]{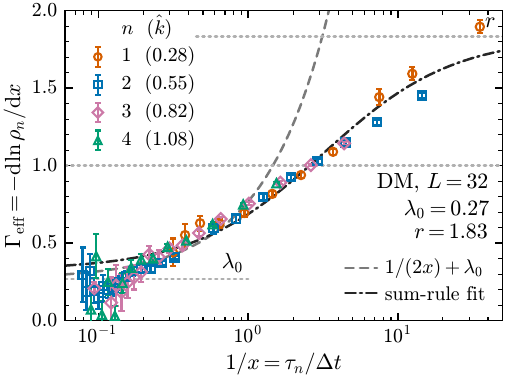}
 \caption{Effective decay rate
 $\Gamma_{\rm eff}=-d\ln\rho_n/dx$ of the DM scaling function against
 $1/x=\tau_n/\Delta t$ for the data of \Fig{fig:critical_scaling}.
 Open markers show the four resolved diagonal modes. The dashed curve
 is a separate tail fit $1/(2x)+\lambda_0$ for $x\geq2$.
 The dash-dotted curve is the logarithmic derivative of the fit in
 \Eq{eq:critical_fit_form}. The dotted lines mark a pure exponential
 decay, $\Gamma_{\rm eff}=1$, the short-time rate $r$ fixed by the
 static amplitudes through \Eq{eq:initial_slope}, and the long-time rate
 $\lambda_0$ from the tail fit.}
 \label{fig:gamma_invx_L32}
\end{figure}

\subsection{Relaxation times and dynamic critical exponent}

The relaxation time of each mode is extracted from the decay of $\rho_n$
with the time separation $\Delta t$. Correlations are measured on two time
grids with spacings $\delta t=0.5$ and $10$. The fine grid resolves the short-time
decay up to $t=50$, while the coarse grid extends the correlation to
$t=1000$.

We obtain $\tau_n$ by linear interpolation between the two neighboring
grid points that bracket the $e^{-1}$ crossing. Dynamic critical scaling predicts
$\tau_n\propto \hat k_n^{-z}$, \Eq{eq:rho_scaling}. We therefore extract the
dynamic critical exponent from the unweighted fit
\begin{equation}
 \ln\tau_n=c-z\ln\hat k_n.
 \label{eq:z_fit}
\end{equation}
At $L=32$, the fit includes $n=1,\ldots,4$. At $L=64$, we use
$n=2,\ldots,8$, where the slowest mode is excluded because
$\tau_1\simeq3\times10^3$ lies beyond the simulated time window.

Statistical uncertainties are estimated from $200$ bootstrap resamples of
whole trajectories. For each dynamics, the same trajectory resample is
used on both time grids, preserving their statistical correlation. Each
bootstrap sample recomputes $\rho_n$, $\tau_n$, and $z$. The quoted
uncertainties are the standard deviations over the bootstrap ensemble.

\subsection{Dynamic scaling function}
\label{app:dynamic_scaling_function}

We fit \Eq{eq:critical_fit_form} jointly to the four resolved modes in
\Fig{fig:critical_scaling}. The fits use the pointwise variances and
do not include correlations between time separations.

\begin{figure}[t]
 \centering
 \includegraphics{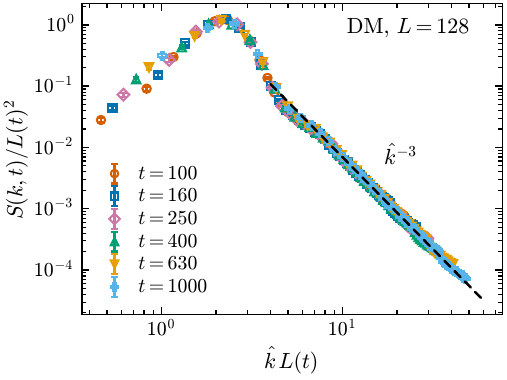}
 \caption{Dynamic scaling function of the structure factor during coarsening.
 Structure factor of the sign field for the $L=128$ DM trajectories of
 \Fig{fig:collapse_L128}, rescaled by the domain size $L(t)$ as in
 \Eq{eq:sk_scaling}. The dashed line is the Porod law $\hat k^{-3}$ with
 fitted amplitude. Error bars are standard errors over the $32$
 trajectories.}
 \label{fig:sk_collapse_L128}
\end{figure}

Previous studies describe long-wavelength relaxation using
near-Lorentzian critical spectra~\cite{Schweitzer2022,Roth2023},
approximate dynamical eigenmodes of Kawasaki
dynamics~\cite{Kawasaki1966,Zhong2019}, and exponential fits in
three-dimensional Model B~\cite{Chattopadhyay2023}.
\Eq{eq:critical_fit_form} allows an algebraic correction to an exponential
tail, $f(x)\sim (ax)^{-b}e^{-\lambda_0x}$ at large $x$.

The short-time behavior is constrained independently by the equilibrium
dynamics. For Model B, conservation and the
fluctuation-dissipation relation give~\cite{HohenbergHalperin1977}
\begin{equation}
 -\rho_n'(0^+)=\frac{\hat k_n^2}{G_n}.
 \label{eq:initial_slope}
\end{equation}
After rescaling time by $\tau_n$, the initial decay rate is therefore
$\tau_n\hat k_n^2/G_n$. Dynamic scaling requires this quantity to become mode
independent. We define $r$ as the inverse-variance weighted mean of
$\tau_n\hat k_n^2/G_n$ over the four modes and impose
\begin{equation}
 \lambda_0+ab=r
 \label{eq:slope_constraint}
\end{equation}
in the joint fits of \Eq{eq:critical_fit_form} shown in
\Fig{fig:critical_scaling}. Fits with all three parameters free yield
initial rates within a few percent of $r$ for both dynamics.
This consistency with the static fluctuations was not imposed during
training.

\begin{figure}[t]
 \centering
 \includegraphics[width=\columnwidth]{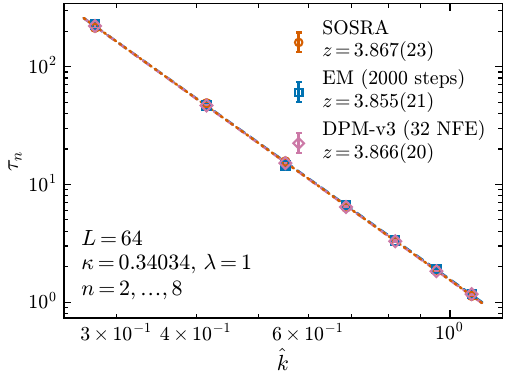}
 \caption{Reverse-sampling resolution at $L=64$.
 Relaxation times are extracted by joining the statistically resolved segments
 of time-origin-averaged correlations on output grids with spacings
 $\delta t=0.5$ up to $t=50$ and $\delta t=10$ up to $t=1000$. We compare SOSRA,
 the learned kernels sampled with 2000 EM steps, and the same kernels
 sampled with DPM-Solver-v3 using 32 NFE per physical transition.
 Only the commonly resolved modes $n=2,\ldots,8$ are shown. The dotted line
 for SOSRA and dashed lines for the learned kernels are power-law fits to
 $\tau_n\propto \hat k_n^{-z}$. Error bars and the parenthesized uncertainties
 of $z$ are trajectory-bootstrap standard errors.}
 \label{fig:reverse_nfe_tau_L64}
\end{figure}

We also examine the effective decay rate
\begin{equation}
 \Gamma_{\rm eff}(x)=-\frac{d\ln\rho_n}{dx}.
 \label{eq:gamma_eff}
\end{equation}
For the fit in \Eq{eq:critical_fit_form},
\begin{equation}
 \Gamma_{\rm eff}(x)=\lambda_0+\frac{ab}{1+ax}.
 \label{eq:gamma_fit}
\end{equation}
\Fig{fig:gamma_invx_L32} estimates this rate directly from the
correlations. Each point is the logarithmic
difference of $\rho_n$ between two collapsed points on the same time grid,
separated by a fraction of $x$ and placed at their midpoint. Error bars
are bootstrap standard errors of the difference over trajectories at
fixed $\tau_n$. The dash-dotted curve,
labeled sum-rule fit, is \Eq{eq:gamma_fit} with the parameters of the
joint fit of \Eq{eq:critical_fit_form} under the constraint
\Eq{eq:slope_constraint}. The dashed curve is a separate fit of
$\lambda_0+1/(2x)$ to the points with $x\geq2$, with only $\lambda_0$
free. It is the large-$x$ form of \Eq{eq:gamma_fit} with $b=1/2$.
The fits describe an approximate crossover over the measured range,
with parameters that depend on the mode and fitting window.

\subsection{Coarsening structure factor}
\label{app:coarsening_sk}

\begin{figure*}[t]
 \centering
 \includegraphics[width=\textwidth]{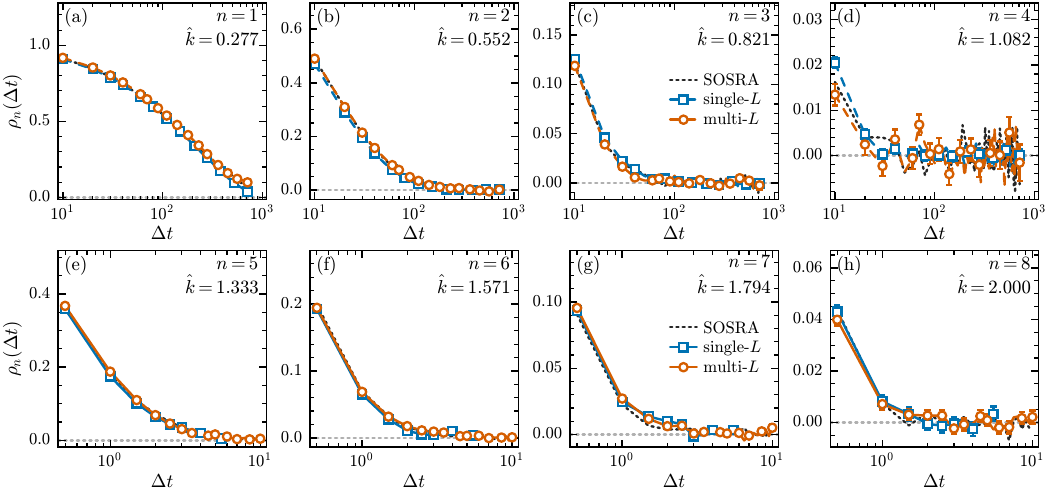}
 \caption{Time-origin-averaged diagonal-mode correlations
 $\rho_n(\Delta t)$ on the equilibrium ensemble at $L=32$. Panels
 (a)--(d) show $n=1,\ldots,4$ for kernels trained at $L=32$ and at
 $L=8,16,32$, with training lag $\delta t=10$.
 Panels (e)--(h) show $n=5,\ldots,8$ for both kernels with
 $\delta t=0.5$. Dotted lines without markers show SOSRA on the corresponding
 time grid, squares the kernel trained at $L=32$, and circles the kernel
 trained at $L=8,16,32$. For the kernels, dashed and solid lines denote
 $\delta t=10$ and $0.5$.
 The upper row is shown up to $\Delta t=710$. In the lower row, the kernel
 trained at $L=32$ is shown up to $\Delta t=5.5$, and the other curves up to
 $\Delta t=10$. Every point averages at least 30 time origins.
 Error bars are leave-one-trajectory-out jackknife standard errors
 over 512 trajectories for training at $L=8,16,32$ and over 1024 trajectories
 for training at $L=32$.}
 \label{fig:L32_chain}
\end{figure*}

The equal-time structure factor of the sign field is the Fourier
transform of the site-averaged correlation in \Eq{eq:sign_correlation},
taken before the average over orientations of $r$,
\begin{equation}
 S(k,t)=\frac{1}{V}\left\langle|\sigma(k,t)|^2\right\rangle
 =\sum_r e^{-ik\cdot r}\,C(r,t),
 \label{eq:structure_factor}
\end{equation}
with $\sigma(k,t)=\sum_x e^{-ik\cdot x}\sigma(x,t)$. The
ensemble average is taken first. The result is then averaged over
shells of width $2\pi/L$ in the lattice momentum $\hat k$,
and plotted at the mean $\hat k$ of
each shell. The real-space correlation
is averaged in the same way over radial lattice bins. Its first zero is
located by linear interpolation and defines $L(t)$, as in the main text.
The scaling form in \Eq{eq:coarsening_scaling} then gives the dynamic
scaling form of the structure
factor~\cite{BinderStauffer1974,Furukawa1985,Bray1994},
\begin{equation}
 S(k,t)=L(t)^{2}\,h\big(kL(t)\big),
 \qquad
 h(x)\propto x^{-3}
 \;\;\text{for}\;\; x\gg1,
 \label{eq:sk_scaling}
\end{equation}
where $h$ is the two-dimensional Fourier transform of $g$. The tail is
the Porod law for sharp domain walls, $S\propto k^{-(d+1)}$ in $d$
dimensions~\cite{Porod1951,Debye1957,Bray1994}. It follows from the
linear short-distance behavior $g(u)=1-au+\dots$ of a two-phase
correlation, where $a$ is proportional to the wall length per unit
area. The sign field takes only the values $\pm1$, so its walls are
steps between neighboring sites whatever the width of the walls in
$\phi$. The Porod law therefore applies from the inverse domain size up
to the lattice cutoff. For the sign field the exponent is fixed by
construction. The nontrivial content is the amplitude, which collapses
only if the wall length per unit area scales as $1/L(t)$. On the
lattice, a step between neighboring sites contributes $1/\hat k_\mu^2$ in
place of $1/k_\mu^2$ for the wave-vector component normal to it. The tail
is therefore compared with $\hat k^{-3}$. Against the continuum wave number $k$ it bends upward
near the zone boundary, and the different times separate there.
The same scaling form and Porod tail have been observed in direct
simulations of conserved dynamics in two
dimensions~\cite{Amar1988,Rogers1988,PuriOono1988}.

\Fig{fig:sk_collapse_L128} shows $S(k,t)/L(t)^{2}$ against $\hat kL(t)$
for the $L=128$ DM ensemble of \Fig{fig:collapse_L128}, at the same times
between $100$ and $1000$. The length $L(t)$ is taken from the real-space
correlation, without a separate fit in momentum space. The rescaled
structure factors collapse onto a single curve with a $\hat k^{-3}$
tail, except for the lowest shells at $\hat kL(t)\lesssim1.5$, whose
wavelengths are comparable to the box size and which deviate by more
than their statistical errors. The collapse and the Porod tail are
consistent with the domain-growth analysis in the main text.

The $t^{1/3}$ regime has also been reported in two-dimensional
Cahn--Hilliard simulations at critical composition~\cite{Konig2021},
in polymer models with diffusive transport~\cite{Garcia2023}, and in
experiments on surface-directed demixing of polymer
mixtures~\cite{Krausch1993}. Related coarsening regimes occur in passive
lipid membranes~\cite{Zhiliakov2021,Arnold2023} and in models of HP1
condensates without chromatin coupling~\cite{Tortora2023}.

\section{Training on multiple sizes and transfer tests}
\label{app:multiL}

\begin{figure*}[t]
 \centering
 \includegraphics{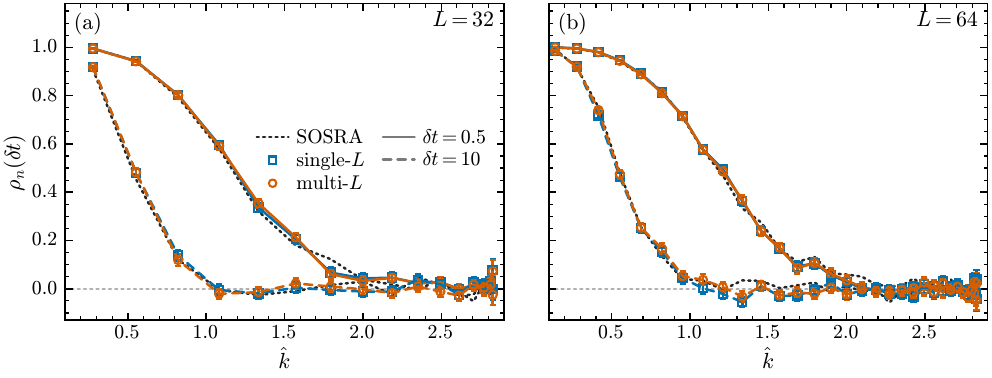}
 \caption{One-step diagonal-mode transfer tests at $L=32$ and $L=64$.
 Panels (a) and (b) show $\rho_n(\delta t)$ at $L=32$ and $64$, respectively.
 Each correlation is evaluated at the single-step separation $\Delta t=\delta t$.
 The $L=32$ initial fields are drawn from the training set, while $L=64$ was
 excluded from training.
 Each panel compares $\delta t=0.5$ and $10$ for SOSRA and the two diffusion
 kernels. Dotted lines without markers show SOSRA at both time separations.
 For the kernels, solid lines denote $\delta t=0.5$ and dashed lines
 $\delta t=10$. Squares show the kernel trained only at $L=32$, and
 circles show the kernel trained on mixed $L=8,16,32$ lattices.
 Error bars on the kernel results are
 leave-one-trajectory-out jackknife standard errors over 512 trajectories.}
 \label{fig:one_step_comparison}
\end{figure*}

\Fig{fig:reverse_nfe_tau_L64} applies the kernels used in
\Fig{fig:critical_scaling} to $L=64$, beyond the training sizes $L=8,16,32$.
The network uses local convolutions with periodic boundaries and accepts
the larger lattice without changing its weights.

\subsection{Relaxation and reverse sampling}

\Fig{fig:reverse_nfe_tau_L64} compares $512$ trajectories for each dynamics
at $L=64$. Relaxation times are extracted using the two output grids and
the procedure in \App{app:observable_details}.

We also test the dependence on the reverse integrator while keeping the
trained score networks fixed. For the normalized noisy increment $\psi_s$ and
condition $\varphi$, the score gives the noise prediction at level $\sigma_s=\Sigma(s)$,
\begin{equation}
 \epsilon_\theta(\psi_s,\varphi,\sigma_s)
 =-\sigma_s\mathcal P_0s_\theta(\psi_s,\varphi,s).
 \label{eq:score_to_noise}
\end{equation}
This noise coordinate follows the EDM convention~\cite{Karras2022EDM}.
We integrate the corresponding probability-flow ODE with
DPM-Solver-v3~\cite{Zheng2023DPMSolverV3}, following the diffusion-ODE
solvers of Refs.~\cite{Lu2022DPMSolver,Lu2022DPMSolverPP}.
The solver coefficients are estimated from $256$ original $L=32$ training
pairs for each training lag $\delta t$. No $L=64$ test field enters this calculation.
Sampling uses $32$ steps uniform in $\ell=-\log\sigma_s$, requiring $32$
network function evaluations (NFE) per physical transition. The trained
networks and training lags are the same as for EM sampling.
The zero-mode projection is applied to the initial noise and every
predicted noise field.

For the slowest resolved mode in \Fig{fig:reverse_nfe_tau_L64},
SOSRA gives $\tau_2=213.83\pm7.96$, EM with
2000 steps gives $226.82\pm7.61$, and DPM-Solver-v3 gives
$221.28\pm7.29$.  Fits over the common modes $n=2,\ldots,8$ give
$z=3.867\pm0.023$, $3.855\pm0.021$, and $3.866\pm0.020$, respectively.
The fitted exponents agree within the bootstrap errors.

The reduction from 2000 EM steps to 32 NFE produces no statistically
resolved change in the relaxation spectrum.
The residual offset in the equal-time power is insensitive to the two
sampling schemes tested here.

The exponents in \Fig{fig:reverse_nfe_tau_L64} characterize the finite
momentum range of the simulations. The asymptotic Model B relation
$z=4-\eta$~\cite{Halperin:1974zz,Halperin:1976zz} gives $z=15/4=3.75$
in the two-dimensional Ising universality class. Monte Carlo
renormalization group studies found $z=3.80$~\cite{YalabikGunton1982}
and $3.79(5)$~\cite{Moseley1989}. Kawasaki simulations gave
$3.95(10)$ after correction of the update
scheme~\cite{Zheng2000,Zheng2001Erratum}. Classical statistical simulations
of relativistic diffusion found $3.83(10)$ and $3.716(17)$ at different
couplings to the heat bath~\cite{Schweitzer2022}.

In three dimensions, simulations of Model B using conserved Metropolis
updates yielded $z=3.972(2)$ from finite-size
scaling~\cite{Chattopadhyay2023}.

\subsection{Correlations at the training size and one-step transfer}

\Fig{fig:L32_chain} compares the equilibrium $L=32$ correlations from SOSRA
with those from repeated sampling of the learned kernels. The low modes use
$\delta t=10$ and compare training at $L=32$ with training at $L=8,16,32$.
The faster modes use the latter training data with $\delta t=0.5$.

\Fig{fig:one_step_comparison} then compares a single physical transition
at $L=32$ and $64$ for both training choices. The $L=32$ one-step
comparison uses $512$ initial fields from the training set. The
$L=64$ initial fields are absent from training. Both training datasets here
use canonical fields. The kernel in \Fig{fig:validation_L32} uses Gaussian
training fields.

\section{Computational cost}
\label{app:computational_cost}

\begin{table*}[t]
\caption{Accuracy at $t=100$ and wall time for the non-equilibrium benchmark. Observable
entries are signed relative differences from the $h=0.0001$ SOSRA reference
in percent.  The last column is Eq.~\eqref{eq:non-equilibrium_solver_error}.
All rows use one NVIDIA GH200.}
\label{tab:non-equilibrium_solver_accuracy}
\begin{ruledtabular}
\begin{tabular}{lccrrrr}
Method & Step or NFE & Time (s) & $S/V$ & $\phi_2$ & $\phi_4$ & $\epsilon$ \\
\hline
DPM-Solver-v3 (This work) & 32 NFE per 10 & $3.26$ & $-0.078$ & $+0.098$ & $+0.183$ & $0.128$ \\
SOSRA (This work) & $0.0025$ & $3.28$ & $-0.075$ & $-0.061$ & $-0.157$ & $0.107$ \\
SOSRA (StochasticDiffEq.jl)~\cite{Rackauckas2020SOSRA,Rackauckas2017DifferentialEquations} & $0.0025$ & $30.67$ & $-0.146$ & $-0.021$ & $-0.102$ & $0.104$ \\
ShARK (Diffrax)~\cite{Foster2024ShARK,Kidger2021Diffrax} & $0.00125$ & $17.76$ & $+0.007$ & $-0.052$ & $-0.130$ & $0.081$ \\
EM (Diffrax)~\cite{Maruyama1955,Kidger2021Diffrax} & $0.0002$ & $38.75$ & $+0.209$ & $-0.047$ & $-0.025$ & $0.125$ \\
EM (StochasticDiffEq.jl)~\cite{Maruyama1955,Rackauckas2017DifferentialEquations} & $0.0002$ & $256.44$ & $+0.163$ & $-0.005$ & $+0.049$ & $0.098$ \\
EM (torchsde)~\cite{Maruyama1955,Li2020torchsde} & $0.0002$ & $356.89$ & $+0.368$ & $-0.099$ & $-0.076$ & $0.225$ \\
\end{tabular}
\end{ruledtabular}
\end{table*}

Table~\ref{tab:gpu_wall_time} compares the time required to evolve $512$ configurations
on an $L=64$ lattice from $t=0$ to $100$. Both benchmarks use one NVIDIA
GH200. Each reported time is a single synchronized measurement after
warmup. Model loading and Julia, PyTorch, and JAX compilation are excluded.
The direct solvers use single precision. The DPM-Solver-v3 network uses
bfloat16, with single- and double-precision solver arithmetic.

\subsection{Equilibrium benchmark}
\label{app:timing_benchmark}

The equilibrium column of Table~\ref{tab:gpu_wall_time} starts every method from
the same equilibrium initial fields at $\kappa=0.34034$ and $\lambda=1$.
The learned kernel is the $\delta t=10$ kernel trained on canonical pairs used in
\App{app:multiL}, sampled with DPM-Solver-v3 at $32$ NFE per transition.

DPM-Solver-v3 advances ten physical intervals. Both SOSRA implementations
use step size $h=0.00625$. ShARK uses $h=0.004$, and Diffrax EM uses
$h=0.0004$.

The step sizes are chosen to give comparable action-density accuracy after
one physical interval. For the five explicitly timed methods in the order
of Table~\ref{tab:gpu_wall_time}, the shifts at $t=10$ are $+0.719\%$,
$-0.689\%$, $-0.825\%$, $-0.733\%$, and $+0.775\%$.
The completed runs of the other tested EM implementations take more than
$100$ s for the same $t=100$ evolution.

\subsection{non-equilibrium benchmark}
\label{app:non-equilibrium_timing}

The non-equilibrium column of Table~\ref{tab:gpu_wall_time} uses $512$
disordered initial fields of the kind used in \Fig{fig:collapse_L128},
here at $L=64$. A fused SOSRA calculation with $h=0.0001$ supplies the
reference ensemble at $t=100$.

The step sizes were calibrated using the action density and the spatial
moments $\phi_p=V^{-1}\sum_x\phi_x^p$ for $p=2,4$. For method $m$ we define
\begin{equation}
 \epsilon_m=100\left[
 \frac{1}{3}\sum_{O\in\{S/V,\phi_2,\phi_4\}}
 \left(
 \frac{\langle O\rangle_m-\langle O\rangle_{\rm ref}}
 {|\langle O\rangle_{\rm ref}|}
 \right)^2
 \right]^{1/2}.
 \label{eq:non-equilibrium_solver_error}
\end{equation}
DPM-Solver-v3 gives $\epsilon=0.128\%$. For SOSRA, ShARK, and Diffrax EM,
the calibrated steps give errors from $0.081\%$ to $0.125\%$.
The resulting values and the individual signed shifts are listed in
Table~\ref{tab:non-equilibrium_solver_accuracy}.

The StochasticDiffEq.jl and torchsde EM runs use the step calibrated with
Diffrax. Their wall times are $256.44$ s and $356.89$ s, respectively,
giving the $>200$ s entry in Table~\ref{tab:gpu_wall_time}.

Under these accuracy criteria, DPM-Solver-v3 takes $3.24$ s in equilibrium
and $3.26$ s out of equilibrium. These times are below those of the public
implementations tested here. The fused SOSRA implementation is faster in
equilibrium at $1.30$ s and comparable out of equilibrium at $3.28$ s,
as reported in Table~\ref{tab:gpu_wall_time}.

\section{Colloid experiment}
\label{app:experiment}

We test long-time prediction on measured trajectories of colloidal
particles in a periodic optical
potential~\cite{dieball2025precisely,Dieball2024data}. Separate diffusion
models learn the undriven and driven motion from the first $10$ min of
each recording and predict the subsequent $50$ and $20$ min, respectively.
Under driving, the predicted mean speed and mean passage time agree with
the measurements to within $1\%$. The comparisons below examine transport
and transition paths, using the Langevin model of the original study as
a reference.

\begin{table*}[t]
\caption{Measured and predicted colloid transport at $10$ s resolution.
The comparison covers $50$ min without driving and $20$ min under driving,
starting at recording time $10$ min. Parentheses give standard errors over
measured tracks. Model means use $32$ replicas per initial position, with
sampling errors omitted. Passage times include only trajectories reaching
either $x_{\rm i}\pm\lambda$ without driving, or $x_{\rm i}-\lambda$ under
driving, within the observation window.}
\label{tab:experiment}
\begin{ruledtabular}
\begin{tabular}{lcccccc}
 & \multicolumn{3}{c}{Undriven} & \multicolumn{3}{c}{Driven} \\
\cline{2-4}\cline{5-7}
 & Data & DM & Langevin & Data & DM & Langevin \\
\hline
$P(\Delta n=-1)$ per step & $0.0157$ & $0.0154$ & $0.0161$ & $0.0728$ & $0.0735$ & $0.0763$ \\
$P(\Delta n=+1)$ per step & $0.0168$ & $0.0170$ & $0.0161$ & $0.0047$ & $0.0047$ & $0.0043$ \\
$\bar v$ [$\mu$m/s] & $0.0005(1)$ & $0.0006$ & $0.0000$ & $-0.0282(3)$ & $-0.0284$ & $-0.0298$ \\
$D_{{\rm eff},x}$ [$\mu$m$^2$/s] & $0.0152$ & $0.0163$ & $0.0163$ & $0.0430$ & $0.0396$ & $0.0408$ \\
Passage time [s] & $610(15)$ & $576$ & $576$ & $168(4)$ & $166.5$ & $157.3$ \\
\end{tabular}
\end{ruledtabular}
\end{table*}

\begin{figure*}[t]
 \centering
 \includegraphics{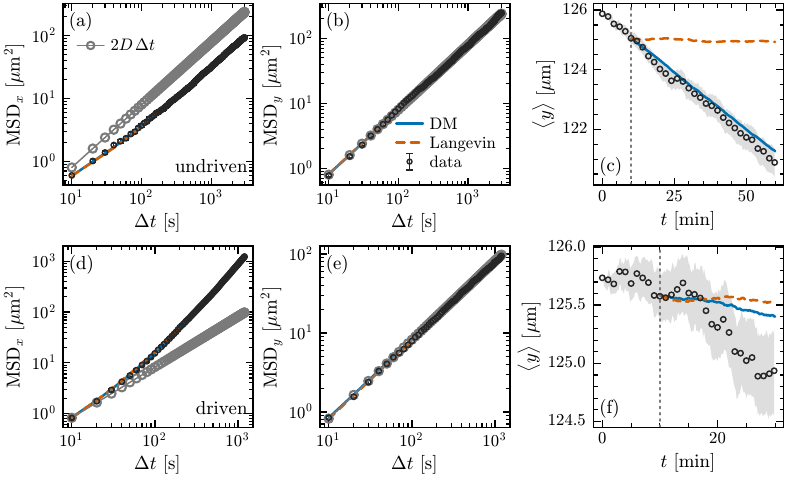}
 \caption{Colloid transport without driving (upper row) and under driving
 (lower row). Symbols and lines as in \Fig{fig:experiment}.
 (a,b,d,e)~Mean-square displacement along $x$ and $y$ over the time
 $\Delta t$ elapsed since the common origin at $t=10$ min. The gray line
 shows free diffusion $2D\,\Delta t$.
 (c,f)~Mean transverse position against recording time. Shading shows
 the standard error of the mean displacement from $t=0$. The dotted line
 marks the end of training. The later transverse drift in the driven
 data is absent from both models.}
 \label{fig:experiment_transport}
\end{figure*}

\begin{figure*}[t]
 \centering
 \includegraphics[width=\textwidth]{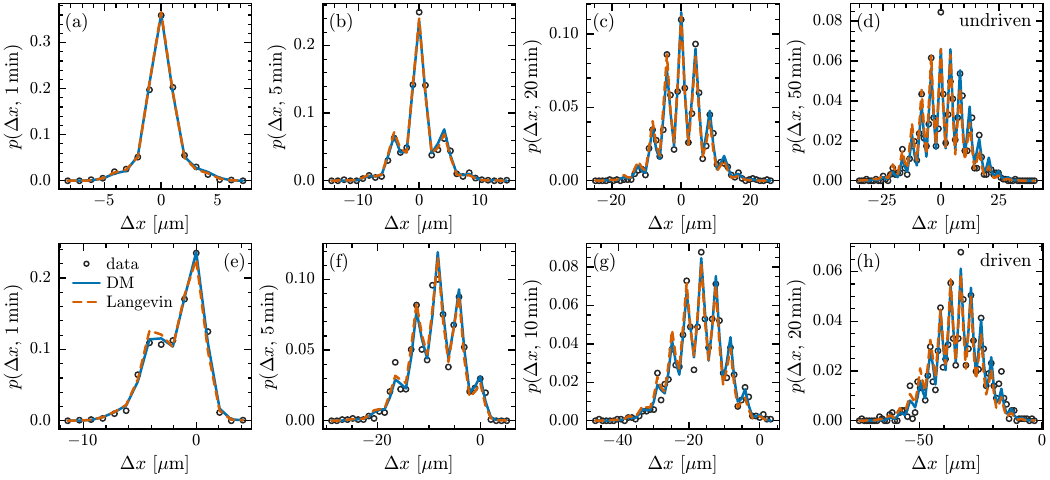}
 \caption{Distributions of the net displacement $\Delta x$ from $t=10$ min
 after $1$, $5$, $20$, and $50$ min of undriven motion (upper row) and
 after $1$, $5$, $10$, and $20$ min of driven motion (lower row). Panels
 (d) and (h) repeat \Fig{fig:experiment}(b,f). Bins have width $\lambda/4$
 and are centered at integer multiples of $\lambda/4$. Symbols and lines
 as in \Fig{fig:experiment}.}
 \label{fig:experiment_pdx_time}
\end{figure*}

\subsection{Experimental system and observations}

The trajectories are the public dataset~\cite{Dieball2024data} of
Ref.~\cite{dieball2025precisely}. Micron-sized polystyrene particles
diffuse near the bottom plate of a quasi-two-dimensional cell, and
interfering laser beams imprint a periodic potential along $x$ with period
$\lambda=4.1\,\mu$m. About $1200$ complete tracks are available in each
case, $60$ min long without driving and $30$ min long under driving. In
the driven experiment the cell moves relative to the fixed
optical pattern at $v_0=0.053\,\mu$m/s. Recordings with the opposite
direction are reflected, so the mean driven motion is always toward
negative $x$.

Thermal motion explores wells about $3\,k_BT$ deep. Relaxation within a
well takes about $10$ s, while undriven escape takes about $10$ min.
The training lag $\delta t=10$ s therefore resolves motion within a well, and many
successive steps are needed to predict escape and transport. Driving
biases the barrier crossings and produces a net particle current.

We represent the initial position by its phase in the periodic potential.
For each recording, the period $\lambda$ and fringe tilt $b$ are fitted
using only the first $10$ min, by maximizing
$|\langle\exp[2\pi i(x/\lambda-by)]\rangle|$. The phase of this average
sets the origin $x_0$, giving
\begin{equation}
 u=\frac{x-x_0}{\lambda}-by.
 \label{eq:colloid_phase}
\end{equation}
This convention aligns the periodic density across recordings. Under
driving, the density is shifted relative to the potential, so $u=0$ need
not mark a potential minimum. All phase parameters remain fixed during
the subsequent predictions.

\subsection{Dynamics learned from short trajectories}

Each diffusion model learns the displacement
$\Delta r=r(t+\delta t)-r(t)$, with $r=(x,y)$, conditioned on
$c=(\cos2\pi u,\sin2\pi u)$. This imposes spatial periodicity. The
potential, diffusion coefficient, and driving force are not supplied.
Both components of the displacement are learned from the measured pairs.

The forward process and loss are those of
\App{app:conditional_construction}, with a fully connected score network
in place of the convolutional one. Training samples tracks and initial
frames uniformly within the first $10$ min. The reverse SDE is integrated
as in Model B.

Predictions start from each measured position at $t=10$ min, with $32$
independent realizations per particle. The kernel is applied $300$ times
without driving and $120$ times under driving to cover the remaining
recording. The phase is updated after every step, while positions remain
unfolded and can extend beyond the camera's field of view.

\subsection{Langevin reference}

The reference is the overdamped Langevin model of
Ref.~\cite{dieball2025precisely}, expressed in the reflected coordinates,
\begin{align}
 dx&=\left[-\frac{D}{k_BT}\partial_x U+v_{\rm d}\right]dt
       +\sqrt{2D}\,dW_x,
 \nonumber\\
 dy&=\sqrt{2D}\,dW_y.
 \label{eq:colloid_langevin}
\end{align}
Here $D=0.04\,\mu$m$^2$/s is the diffusion coefficient measured between
successive frames, and $v_{\rm d}=0$ without driving or $-v_0$ under
driving. The model assumes isotropic white noise and a periodic potential
that is unchanged by the drive. The potential is inferred from the
undriven training density through $U/k_BT=-\ln p_{\rm eq}(u)$.

For driven motion, the potential is translated so that the predicted
density maximum lies at $u=0$, matching the phase convention of the data.
All parameters are fixed before comparison with the later observations.
The reference uses the same initial positions and number of realizations
as DM.

\begin{figure*}[t]
 \centering
 \includegraphics{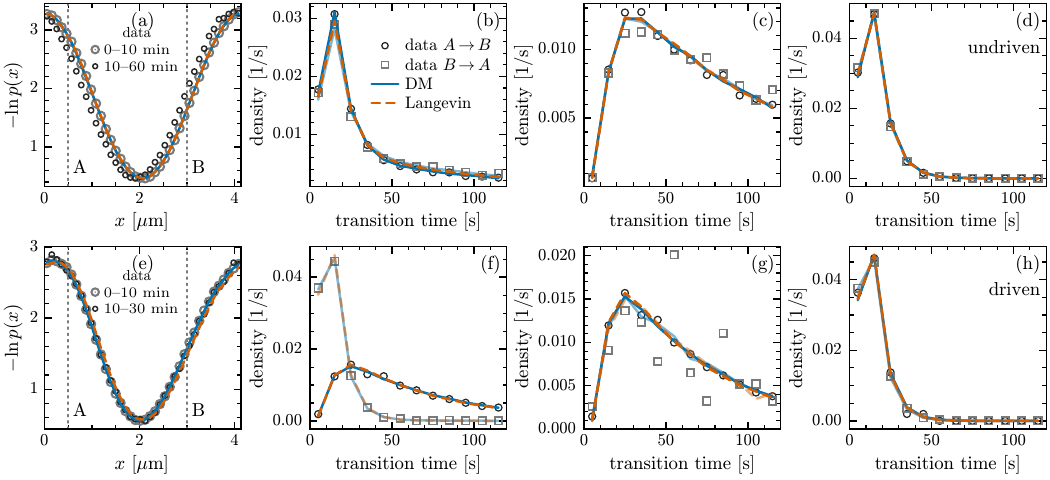}
 \caption{Periodic density and transition-path times of the colloid
 trajectories, undriven (upper row) and driven (lower row).
 (a,e)~$-\ln p(x)$, the negative logarithm of the position probability
 density folded onto one spatial period. The coordinate $x$ measures
 distance from the barrier top along the drive direction. Gray open
 circles show experimental data from the training window ($t=0$--$10$ min).
 Black open circles show data from the subsequent prediction window
 ($t=10$--$60$ min in (a) and $t=10$--$30$ min in (e)). Blue solid and
 orange dashed lines show the DM and Langevin predictions over the
 corresponding prediction window. At equilibrium, $-\ln p(x)$ equals
 $U(x)/(k_BT)$ up to an additive constant. This identification does not
 hold under driving. Dotted lines mark the milestones $A$ and $B$ at $0.5$ and
 $3.0\,\mu$m. (b--d,f--h)~Transition-path times between $A$ and $B$ at
 $10$ s resolution. Black open circles and gray open squares show the data
 for $A\to B$ and $B\to A$. Full and faint lines show DM and the Langevin
 reference for the two directions. (b,f)~All paths. (c,g)~Paths passing
 the well bottom between the milestones. (d,h)~Paths passing the barrier
 top. Without driving, $56\%$ of the $A\to B$ paths and $60\%$ of the
 $B\to A$ paths pass the well bottom. Under driving these fractions are
 $99\%$ and $2\%$.}
 \label{fig:experiment_paths}
\end{figure*}

\subsection{Observables and uncertainty}

The data, DM, and Langevin trajectories are analyzed by the same procedure
on a common time grid with spacing $\delta t=10$ s, matching the training lag.
Displacements are measured from
the common origin at $t=10$ min. Over the observation duration
$T_{\rm obs}$, the tabulated mean velocity and effective diffusion
coefficient are
\begin{equation}
 \bar v=\frac{\langle\Delta x\rangle}{T_{\rm obs}},
 \qquad D_{{\rm eff},x}
 =\frac{\operatorname{Var}(\Delta x)}{2T_{\rm obs}}.
 \label{eq:colloid_transport}
\end{equation}
These quantities describe transport over the stated observation window.
The mean-position bands in \Fig{fig:experiment}(a,e) and
\Fig{fig:experiment_transport}(c,f) are standard errors of displacements
measured from the beginning of each recording.

The phase-resolved drift in \Fig{fig:experiment}(c,g) is
\begin{equation}
 v_{\Delta t}(u_0)
 =\frac{\langle x(t+\Delta t)-x(t)\mid u(t)=u_0\rangle}{\Delta t},
 \label{eq:colloid_finite_time_drift}
\end{equation}
evaluated at $\Delta t=\delta t=10$ s using pairs after the training window
and binned by initial phase. The gray curves show the instantaneous drift
$v_{\rm drift}=-D\partial_xU/(k_BT)+v_{\rm d}$ of
\Eq{eq:colloid_langevin}. The measured $v_{\Delta t}$ averages the drift
along the paths explored during $10$ s, which reduces its variation with
initial phase. Data and model predictions are therefore compared at the
same finite lag.

Hopping probabilities count changes of the well index
$n=\lfloor u+1/2\rfloor$ over single steps $\delta t$ after the training window.
For an initial position $x_{\rm i}$, the undriven passage time is the first
exit from $[x_{\rm i}-\lambda,x_{\rm i}+\lambda]$. Under driving it is the
first arrival at $x_{\rm i}-\lambda$. Crossings are detected on the sampled
time grid. \Fig{fig:experiment}(d,h) shows the mean passage time among
trajectories that cross before a variable cutoff.
Table~\ref{tab:experiment} uses the full observation window, within which
nearly all trajectories cross. The gray line in \Fig{fig:experiment}(h)
estimates the time to travel one period from the mean speed during
training. Sampling every $10$ s delays the detection of crossings and can
miss brief excursions, so all comparisons use the same time grid.

Transition-path times use the milestones of
Ref.~\cite{dieball2025precisely}, two points $0.5$ and $3.0\,\mu$m from
the barrier top along the drive [\Fig{fig:experiment_paths}(a,e)]. A path
runs from the last crossing of one milestone to the first crossing of the
other. The two routes pass either the well bottom between the milestones
or the barrier top on the other side of the period. Returns within one
sampling interval are unresolved, making the measured path times longer
than those obtained at the original frame rate. At full resolution, the
undriven times and route fractions are close to those reported in
Ref.~\cite{dieball2025precisely}.

\subsection{Transport and transition paths}

The displacement distributions in \Fig{fig:experiment_pdx_time} retain
peaks separated by the potential period as they broaden with time.
Under driving, their envelope moves with the particle current. DM
captures both features, reproducing localization within the wells
together with transport across many periods.

\emph{Undriven motion.} A weak current persists without imposed driving
[\Fig{fig:experiment}(a)]. DM learns this bias from the initial
observations, while the equilibrium Langevin reference has zero mean
current. The measured drift decreases during the recording, so the DM
mean position eventually exceeds the data. DM also captures the
directional asymmetry of the transition paths, including the greater
weight of the well-bottom route for $B\to A$
[\Fig{fig:experiment_paths}(b)]. The Langevin reference gives symmetric
path statistics.

Both models predict a mean exit time of about $576$ s, compared with
$610(15)$ s in the data, and an effective diffusion coefficient about
$7\%$ too large (Table~\ref{tab:experiment}). Agreement is closer for the
single-step hopping probabilities and short-time diffusion. The origin
of the remaining long-time discrepancy is not resolved by these
comparisons.

\emph{Driven motion.} The measured mean velocity is
$-0.0282(3)\,\mu$m/s, about half the cell speed. DM gives
$-0.0284\,\mu$m/s and a mean passage time of $166.5$ s, compared with the
measured $168(4)$ s. Both predictions agree to within $1\%$. The Langevin
values are $-0.0298\,\mu$m/s and $157.3$ s, differing by about $6\%$
[\Fig{fig:experiment}(e,h)]. The learned dynamics thus predicts the
driven current and passage time directly from observed motion, without
inferring a potential from the undriven density.

Driving also changes which routes connect the milestones. About $99\%$
of $A\to B$ paths pass the well bottom, compared with only $2\%$ of
$B\to A$ paths. The mean path times in the two directions consequently
differ by a factor of about five. Once paths are separated by route,
the forward and backward time distributions are similar in the data and
both models [\Fig{fig:experiment_paths}(g,h)]. The directional asymmetry
of the combined distributions is therefore accounted for by their
different route weights.

Both models capture the increase in dispersion under driving, although
their effective diffusion coefficients remain below the measured value
(Table~\ref{tab:experiment}). The data also show changes during the
recordings. The undriven drift decreases, the driven
$\operatorname{Var}(\Delta x)/\Delta t$ continues to grow at late times, and a
transverse drift appears only after the training window
[\Fig{fig:experiment_transport}(f)]. The learned kernel and the Langevin
reference do not reproduce this later transverse drift. Their fixed
parameters contain no description of changes in the experimental
conditions.

\emph{Dependence on the time lag.} Halving the training lag to
$\delta t=5$ s leaves the overall comparison similar. The predicted
driven mean speed changes by less than $1\%$, and passage-time deviations
remain comparable when data and models are evaluated at the same
resolution. For both lags and both ensembles, single-step predictions
from training positions reproduce the displacement width and hopping
probability to within $2\%$. These checks support the stability of the
transport predictions under a change of lag, while the long-time
comparison retains the discrepancies described above.

\end{document}